\documentclass[12pt]{article}

\usepackage{graphicx}

\def\la{\mathrel{\mathpalette\fun <}}

\def\fun#1#2{\lower3.6pt\vbox{\baselineskip0pt\lineskip.9pt
\ialign{$\mathsurround=0pt#1\hfil##\hfil$\crcr#2\crcr\sim\crcr}}}

\begin{document}

\title{\bf Charmed penguin versus BAU}
\author{A.D.~Dolgov \\
\small{\em A.I.~Alikhanov Institute of Theoretical and Experimental Physics, Moscow} \\ 
\small{\em Novosibirsk State University, Novosibirsk, 630090, Russia}\\
\small{\em Dipartimento di Fisica, Universit\`a degli Studi di Ferrara, I-44100 Ferrara, Italy} \\
\small{\em Istituto Nazionale di Fisica Nucleare, Sezione di Ferrara,
I-44100 Ferrara, Italy} \\
S.I.~Godunov \\
\small{\em A.I.~Alikhanov Institute of Theoretical and Experimental Physics, Moscow} \\ 
\small{\em Novosibirsk State University, Novosibirsk, 630090,
  Russia}\\
\small{\em VNIIA, Moscow}\\
A.N.~Rozanov \\
\small{\em CPPM IN2P3-CNRS-Universite de Mediterranee, Marseille, France} \\
M.~I.~Vysotsky \\
\small{\em A.I.~Alikhanov Institute of Theoretical and Experimental Physics, Moscow}\\
\small{\em Novosibirsk State University, Novosibirsk, 630090,
  Russia}\\
\small{\em VNIIA, Moscow}}
\date{}
\maketitle

\begin{abstract}
Since  the Standard Model most probably cannot explain the large value of
CP asymmetries recently observed in $D$-meson decays we propose the 
fourth quark-lepton generation explanation of it. As a byproduct weakly 
mixed leptons of the fourth generation make it possible to save the baryon number of 
the Universe from erasure by sphalerons. An impact of the 4th generation on
BBN is briefly discussed.
\end{abstract}

\section{Introduction}

Recently LHCb collaboration has measured the unexpectedly large CP
violating asymmetries in $D\to\pi^+ \pi^-$ and $D\to K^+ K^-$
decays \cite{1}:
\begin{equation}
\Delta A_{CP}^{LHCb} \equiv A_{CP}(K^+ K^-) - A_{CP}(\pi^+ \pi^-)
= [-0.82 \pm 0.21 (\rm{stat.}) \pm 0.11 (\rm{syst.})]\% \;\; ,
\label{1}
\end{equation}
where
\begin{equation}
A_{CP}(\pi^+ \pi^-) = \frac{\Gamma(D^0 \to \pi^+ \pi^-)
-\Gamma(\bar D^0 \to \pi^+ \pi^-)}{\Gamma(D^0 \to \pi^+ \pi^-)
+\Gamma(\bar D^0 \to \pi^+ \pi^-)} \label{2}
\end{equation}
and $A_{CP}(K^+ K^-)$ is defined analogously.

This result was later confirmed by CDF collaboration, which
obtained~\cite{2}:
\begin{equation}
\Delta A_{CP}^{CDF} = [-0.62 \pm 0.21 (\rm{stat.}) \pm 0.10
(\rm{syst.})]\% \;\; . \label{3}
\end{equation}

The most important question concerning experimental results
(\ref{1}) and (\ref{3}) is whether in the Standard Model the CP-violation
(CPV) in these decays can be as large as 0.5\% - 1\%.

In the Standard Model the CPV in $D(\bar D) \to\pi^+ \pi^-$ decays
originates from the interference of the tree and penguin diagrams
shown in Fig. 1. For $D(\bar D)\to K^+ K^-$ decays $d$-quarks in
these diagrams should be substituted by $s$-quarks.

\begin{figure}

\begin{center}

\includegraphics[scale=0.6]{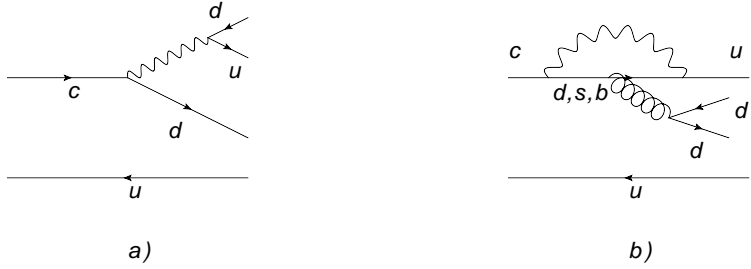}

\caption{Quark diagrams describing $D\longrightarrow\pi^+ \pi^-$
decay in the Standard Model. A wavy line denotes $W$-boson, a
curly line -- gluon.}
\end{center}

\end{figure}

It is convenient to present the penguin diagram contribution to
$D\to\pi^+ \pi^-$ decay amplitude in the following form
\cite{222}:
\begin{eqnarray}
V_{cd}V_{ud}^* f(m_d) + V_{cs}V_{us}^* f(m_s) + V_{cb} V_{ub}^*
f(m_b) = \nonumber \\ = V_{cd} V_{ud}^*[f(m_d)-f(m_s)] + V_{cb} V_{ub}^* [f(m_b) -
f(m_s)] \;\; , \label{4}
\end{eqnarray}
attributing the first term to the tree amplitude and considering
the second term only as the penguin amplitude.

In the case of $D\to K^+ K^-$ decay the following presentation is
useful \cite{222}:
\begin{eqnarray}
V_{cd}V_{ud}^* f(m_d) + V_{cs}V_{us}^* f(m_s) + V_{cb} V_{ub}^*
f(m_b) = \nonumber \\ = V_{cs} V_{us}^*[f(m_s)-f(m_d)] + V_{cb}
V_{ub}^* [f(m_b) - f(m_d)] \, ,
\label{5}
\end{eqnarray}
where the first term is attributed to the tree amplitude while the
second one is the penguin amplitude.

Denoting the absolute values of $D\to\pi^+\pi^-$ decay
amplitudes by $T$ and $P$ we get:
\begin{eqnarray}
A_{\pi^+\pi^-} & = & T\left[1+\frac{P}{T}
e^{i(\delta-\gamma)}\right]
\;\; , \nonumber \\
\bar A_{\pi^+\pi^-} & = & T\left[1+\frac{P}{T}
e^{i(\delta+\gamma)}\right] \;\; , \label{6}
\end{eqnarray}
where $\delta$ stands for the difference of the strong interaction
phases of the tree and the penguin amplitudes, while $\gamma\approx 70^0$
is the phase of $V_{ub}$ (the product $V_{cd} V_{ud}^*$ as well as
$V_{cb}$ are practically real in the standard parametrization of the
CKM matrix).

From eq.~(\ref{6}) for the CPV asymmetry we obtain:
\begin{equation}
A_{CP}(\pi^+ \pi^-) = 2\frac{P}{T} \sin\delta \sin\gamma \;\; ,
\label{7}
\end{equation}
where in the denominator of (\ref{2}) we neglect the terms of the
order of $P/T$ and $(P/T)^2$ which is a very good approximation
because $P/T\sim |V_{cb}V_{ub}^*|/V_{cd} \ll 1$. Here
$\sin\gamma$ is close to unity and we use this value in what follows.

Let us present an argument demonstrating that $\delta$ can also  be
close to 90$^0$. The tree diagram gives dominant contribution to the
$D\to\pi\pi$ decay rates. The corresponding to it 4-fermion Hamiltonian has parts
with isospin 1/2 and 3/2. That is why the produced $\pi$-meson may
have isospin zero or two. So three decay probabilities, $D^+ \to
\pi^+ \pi^0$, $D^0\to\pi^+ \pi^-$, and $D^0 \to \pi^0\pi^0$, depend
on the absolute values of the decay amplitudes  $A_0$ and $A_2$ and their
strong phases difference $\delta_0 -\delta_2$. From the
experimentally measured branching ratios \cite{3}:
$$
{\rm Br}(D^+ \to\pi^+ \pi^0) = [12.6\pm 0.9]\cdot 10^{-4} \; ,
\;\; {\rm Br}(D^0 \to\pi^0 \pi^0) = [8.0\pm 0.8]\cdot 10^{-4} \; ,
$$
\begin{equation}
{\rm Br}(D^0 \to\pi^+ \pi^-) = [13.97\pm 0.26]\cdot 10^{-4}
\label{8}
\end{equation}
we find  for the phase difference of the amplitudes with $I=0$ and $I=2$:
\begin{equation}
|\delta_0 -\delta_2| = 86^0 \pm 4^0 \;\; . \label{9}
\end{equation}
In eq. (\ref{7}) $\delta$ stands for the difference of the
strong phases of  penguin amplitude which has $I=1/2$ and produces
pions with $I=0$ and tree amplitude, which has parts with $I=1/2$
and $I=3/2$ and produces pions with $I=0$ and $I=2$, that is why
$\delta \neq \delta_0 -\delta_2$. Nevertheless eq. (\ref{9})
demonstrates that $\delta$ can be large, and so we
substitute $\sin\delta =1$ into eq. (\ref{7}).

In the limit of $U$-spin ($d\leftrightarrow s$ interchange)
symmetry the tree amplitude of $D(\bar D)\to K^+ K^-$ decay differs by
sign from that of $D(\bar D) \to\pi^+ \pi^-$ decay, while the
penguin amplitudes of these decays are equal, that is why
\begin{equation}
A_{CP}(K^+ K^-) = -A_{CP}(\pi^+\pi^-) \;\; . \label{10}
\end{equation}
However since \cite{3}
\begin{equation}
{\rm Br}(D^0 \to K^+ K^-) = [39.4\pm 0.7] \cdot 10^{-4} \label{11}\,,
\end{equation}
we obtain from eq.~(\ref{8}) that $|A_{K^+ K^-}/A_{\pi^+\pi^-}| \simeq
1.7$ and $U$-spin symmetry is heavily broken in $D$ decays.
Nevertheless let us suppose that (\ref{10}) is not badly violated,
so finally we get:
\begin{equation}
\Delta A_{CP} = 4\,\frac{P}{T} \label{12}\,.
\end{equation}
Now let us try to understand if in the Standard Model we can
obtain
\begin{equation}
\frac{P}{T} = 1.8 \cdot 10^{-3} \label{13}\,,
\end{equation}
which is needed to reproduce the average value of the LHCb and CDF
results.

\section{\mbox{\boldmath$D\to\pi\pi$}: charmed penguin}

Though the four-fermion quark Hamiltonian responsible for these
decays is known, strong interactions does not allow to make an exact 
calculation of the
decay amplitudes. What can be done is an estimate of the decay
amplitudes with the help of factorization. Let us start from the
tree diagram shown in Fig. 1a which dominates in the decay
amplitude:
\begin{eqnarray}
T & = & \frac{G_F}{\sqrt 2} V_{cd} < \pi^+ \pi^-|\bar d
\gamma_\alpha (1+\gamma_5)c\bar u \gamma_\alpha (1+\gamma_5)d|D^0
>
\times \nonumber \\
& \times & \left\{\frac{2}{3}[\alpha_s(m_c)/\alpha_s(M_W)]^{-2/b}
+ \frac{1}{3}[\alpha_s(m_c)/\alpha_s(M_W)]^{4/b}\right\} \;\; ,
\label{14}
\end{eqnarray}
where the last factor originates from the summation of the gluon
exchanges in the leading logarithmic approximation. Substituting
into  it $b=11-2/3 N_f = 23/3$, $\alpha_s(M_W) = 0.12$,
$\alpha_s(m_c) = 0.3$ we find that the factor in the curly brackets
is close to one, $\{ ...  \} = 1.1$. Factorizing the decay amplitude we obtain:
\begin{eqnarray}
T & = & 1.1 \frac{G_F}{\sqrt 2} V_{cd} < \pi^+ |\bar
u\gamma_\alpha (1+\gamma_5)d |0><\pi^- |\bar d
\gamma_\alpha(1+\gamma_5)c| D^0 >=
\nonumber \\
& = & 1.1 \frac{G_F}{\sqrt 2} V_{cd} f_\pi k_{1\alpha}[f_+^\pi(0)
(p+k_2)_\alpha + f_-^\pi(0)(p-k_2)_\alpha] = 1.1 \frac{G_F}{\sqrt
2}V_{cd} f_\pi f_+^\pi(0) m_D^2 \; , \label{15}
\end{eqnarray}
where $k_1$ and $k_2$ are the momenta of the produced
$\pi$-mesons, $p$ is the $D$-meson momentum and we neglect $m_\pi^2$
in comparison with $m_D^2$.

The value of the $D^0\to\pi^+ e^+\nu$ transition formfactor at $q^2=0$ can be 
found in ref.~\cite{3}:
\begin{equation}
f_+^\pi(0)|V_{cd}| = 0.152 \pm 0.005 \; , \;\; f_+^\pi(0) = 0.66
\;\; , \label{16}
\end{equation}
and for the decay width we obtain:
\begin{equation}
\Gamma_{D\to \pi^+ \pi^-}^{\rm theor.} = \frac{G_F^2}{2}
\frac{(1.1 V_{cd} f_+^\pi(0) f_\pi m_D^2)^2}{16\pi m_D} = 6.2
\cdot 10^9 {\rm s}^{-1} \;\; , \label{17}
\end{equation}
where $f_\pi = 130$ MeV was used.

From the branching ratio of the $D^0\to\pi^+ \pi^-$ decay (\ref{8})
and $D^0$-meson mean life, $\tau_{D^0} = 0.41 \cdot 10^{-12}$ s,
we find:
\begin{equation}
\Gamma_{D\to \pi^+ \pi^-}^{\rm exp} = 3.4 \cdot 10^9 {\rm s}^{-1} \;\; , \label{18}
\end{equation}
So the naive factorization overestimates the decay amplitude by
the factor $\sqrt{6.2/3.4}\approx 1.4$.

Calculating the $D\to K^+ K^-$ decay probability we obtain:
\begin{equation}
\Gamma_{D\to K^+ K^-}^{\rm theor} = \left[\frac{f_K}{f_\pi}
\frac{f_+^K(0)}{f_+^\pi(0)}\right]^2 \Gamma_{D\to \pi^+
\pi^-}^{\rm theor} = 12.2 \cdot 10^9 {\rm s}^{-1} \;\; ,
\label{19}
\end{equation}
where we substituted $f_K/f_\pi = 1.27$ and $f_+^K(0) = 0.73$ taken from
ref.~\cite{3}.

From eq.~(\ref{11}) it follows:
\begin{equation}
\Gamma_{D\to K^+ K^-}^{\rm exp} = 9.6 \cdot 10^9 {\rm s}^{-1}
\;\; , \label{20}
\end{equation}
so the factorization overestimates the decay amplitude by the
factor $\sqrt{12.2/9.6} = 1.1$.

We see that in the case of the tree diagrams the accuracy of
the factorization approximation is very good.

Let us make a brief remark on the $D\to K^0 \bar{K^0}$ decay. At the
tree level it proceeds through the diagram with $W$-boson exchange in
$t$-channel, so it should be suppressed. Even more, $c\bar u \to
d\bar d$ and $c\bar u\to s\bar s$ amplitudes interfere
destructively and in the $U$-spin symmetry limit their sum is
zero \cite{4}. According to experimental data \cite{3}:
\begin{equation}
{\rm Br}(D^0 \to K^0\bar{K^{0}}) = 4 {\rm Br}(D^0 \to 2K_{S}^0)= (6.8 \pm 1.2) \cdot 10^{-4} \;\; ,
\label{21}
\end{equation}
which is
approximately 6 times smaller than ${\rm Br}(D\to K^+ K^-)$.
It means that the decay amplitude is smaller than that to charged
kaons by factor 2.5. This unexpectedly small suppression may indicate
that large distance effects like $D^{0}\rightarrow
K^{*+}K^{*-}\rightarrow K^0\bar{K^{0}}$ rescattering can be important.

The four-fermion QCD penguin amplitude which describes $D\to \pi^+
\pi^-$ decay looks like:
\begin{eqnarray}
H(P) & = & \frac{G_F}{\sqrt 2} V_{cb} V_{ub}^*
\frac{\alpha_s(m_c)}{12\pi} \ln \left(\frac{m_b}{m_c}\right)^2
(\bar u\gamma_\alpha(1+\gamma_5) \vec\lambda c)(\bar d
\gamma_\alpha \vec\lambda d) = \nonumber \\
& = & \frac{G_F}{\sqrt 2} V_{cb} V_{ub}^*
\frac{\alpha_s(m_c)}{12\pi} \ln \left(\frac{m_b}{m_c}\right)^2
[(\bar u\gamma_\alpha(1+\gamma_5) d)(\bar d
\gamma_\alpha(1+\gamma_5)c) - \nonumber \\
& - & 2 \bar u(1-\gamma_5)d\bar d(1+\gamma_5)c] \frac{8}{9} \;\; ,
\label{22}
\end{eqnarray}
where $\vec\lambda$ are the Gell-Mann SU(3) matrices and we use
the Fierz identities: 
\begin{eqnarray}
\vec\lambda_{ab}\vec\lambda_{cd}&=& -2/3\delta_{ab} \delta_{cd} + 2\delta_{ad}\delta_{bc}\,,
\nonumber \\
\bar\psi\gamma_\alpha(1+\gamma_5)\varphi\bar\chi\gamma_\alpha(1+\gamma_5)\eta
&=& \bar\psi\gamma_\alpha(1+\gamma_5)
\eta\bar\chi\gamma_\alpha(1+\gamma_5)\varphi \,,
\nonumber\\
\bar\psi\gamma_\alpha(1+\gamma_5)\varphi\bar\chi\gamma_\alpha(1-\gamma_5)\eta
&=& -2\bar\psi(1-\gamma_5) \eta\bar\chi(1+\gamma_5)\varphi\,.
\nonumber
\end{eqnarray}
Also the identity $<\pi^+|\bar u_a Od_b|0> = 1/3 \delta_{ab} <
\pi^+|\bar u Od|0>$, where $O \equiv \gamma_\alpha \gamma_5$ or
$\gamma_5$, was used.

Calculating the matrix element in the factorization approximation
with the help of the equations of motion for quark fields we find:
\begin{equation}
P = \frac{G_F}{\sqrt 2} |V_{cb} V_{ub}^*|
\frac{\alpha_s(m_c)}{12\pi} \ln\left(\frac{m_b}{m_c}\right)^2
\frac{8}{9} f_\pi f_+^\pi(0) m_D^2 \left[1+\frac{2m_\pi^2}{m_c(m_u
+ m_d)}\right] \;\; . \label{23}
\end{equation}
Dividing it by the experimental value of the tree amplitude and using
Eq.(\ref{15}) we obtain:
\begin{equation}
P/T = \frac{1.4}{1.1} \frac{8}{9} \frac{|V_{cb}
V_{ub}^*|}{|V_{cd}|} \frac{\alpha_s(m_c)}{12\pi}
\ln\left(\frac{m_b}{m_c}\right)^2 \left[1+\frac{2m_\pi^2}{m_c(m_u
+ m_d)}\right] \;\; . \label{24}
\end{equation}
Substituting $|V_{cd}| = 0.23$, $|V_{ub}| = 3.9 \cdot 10^{-3}$,
$V_{cb} = 41 \cdot 10^{-3}$, $\alpha_s(m_c) = 0.3$, $m_b = 4.5$
GeV, $m_c = 1.3$ GeV, $m_u +m_d = 6$ MeV we come to:
\begin{equation}
P/T \approx 9 \cdot 10^{-5} \;\; . \label{25}
\end{equation}
Comparing it with eq.~(\ref{13}) we see that in order to fit the
experimental data on $\Delta A_{CP}$ the penguin amplitude should be
enhanced by the factor 20 in comparison with what factorization gives.
Concerning the tree amplitudes, we have found in this section that
factorization result differs from the experimental value by the  factor
1.4 in the case of $D\to \pi^+ \pi^-$ decay and by 1.1 in the case of
$D\to K^+ K^-$ decay. In the next two sections we will study how
accurate is the factorization approximation to the penguin amplitudes
in $B$- and $K$-meson decays.

\section{\mbox{\boldmath$B\to \pi K$}: beautiful penguin}

$B_u \to \pi^+ K^0$ decay is described by the penguin amplitude
shown in Fig. 2.
\begin{figure}

\begin{center}

\includegraphics[scale=0.5]{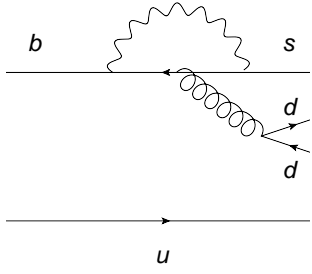}

\caption{$B_u \to \pi^+ K^0$ decay proceeds through the penguin
amplitude only.}

\end{center}
\end{figure}

The Hamiltonian responsible for this decay looks
like:
\begin{equation}
\hat H = \frac{G_F}{\sqrt 2} V_{tb} V_{ts}^* [c_3 O_3 + c_4 O_4 +
c_5 O_5 + c_6 O_6] \;\; , \label{26}
\end{equation}
$V_{tb} V_{ts}^*$ is substituted for $V_{cb}V_{cs}^*+V_{ub}
V_{us}^*$ (the contribution of a loop with the virtual $t$-quark is
negligible) and
\begin{eqnarray}
O_3 & = & \bar s\gamma_\alpha(1+\gamma_5) b\bar d
\gamma_\alpha(1+\gamma_5) d \nonumber \\
O_4 & = & \bar s_a\gamma_\alpha(1+\gamma_5) b_c\bar d_c
\gamma_\alpha(1+\gamma_5) d_a \nonumber \\
O_5 & = & \bar s\gamma_\alpha(1+\gamma_5) b\bar d
\gamma_\alpha(1-\gamma_5) d \nonumber \\
O_6 & = & \bar s_a\gamma_\alpha(1+\gamma_5) b_c\bar d_c
\gamma_\alpha(1-\gamma_5) d_a  \;\; , \label{27}
\end{eqnarray}
where $a,c = 1,2,3$ are the color indexes.

Using the Fierz identities as well as $<K^0|\bar s_a Od_b|0> =
\frac{1}{3}\delta_{ab}< K^0|\bar s Od|0>$ identity we obtain:
\begin{equation}
\hat H = \frac{G_F}{\sqrt 2} V_{tb} V_{ts}^* [a_4 \bar s
\gamma_\alpha(1+\gamma_5) d \bar d \gamma_\alpha(1+\gamma_5) b -
2a_6 \bar s(1-\gamma_5) d\bar d (1+\gamma_5)b] \;\; , \label{28}
\end{equation}
where $a_4 = \frac{1}{3} c_3 + c_4$, $a_6 = \frac{1}{3}c_5 + c_6$.
Calculating the matrix element in the factorization approximation
we obtain:
\begin{equation}
M = \frac{G_F}{\sqrt 2} V_{tb} V_{ts}^* f_K f_+(0) m_B^2 \left[a_4
+ a_6 \frac{2m_K^2}{m_b m_s}\right] \;\; , \label{29}
\end{equation}
where in the leading logarithmic approximation the following
approximate equation is valid:
\begin{equation}
a_4 = a_6 = -\frac{\alpha_s(m_b)}{12\pi}
\ln\left(\frac{M_W}{m_b}\right)^2 \approx -0.03 \;\; , \label{30}
\end{equation}
while at NLO approximation from Table 1 of \cite{5} we obtain:
$a_4 = -0.031$, $a_6 = -0.042$. Substituting $m_s = 100$ MeV, $m_b
= 4.5$ GeV we find:
\begin{equation}
\Gamma_{(B_u \to \pi^+ K^0)}^{\rm fact} =
\frac{G_F^2|V_{ts}|^2}{32\pi}f_K^2 f_+^2(0) m_B^3 (0.076)^2 = 4.1
\cdot 10^6 {\rm s}^{-1} \;\; , \label{31}
\end{equation}
where $V_{ts} = 39\cdot 10^{-3}$ and $f_+(0) = 0.25$ from \cite{3}
was used. The experimental result is:
\begin{equation}
\Gamma_{(B_u \to \pi^+ K^0)}^{\rm exp} = 14 \cdot 10^6 {\rm
s}^{-1} \;\; , \label{32}
\end{equation}
So, the factorization result is enhanced by the factor
\begin{equation}
P/P_{\rm fact} = \sqrt{14/4.1} = 1.8 \;\; . \label{33}
\end{equation}

The numerical value of the penguin amplitude is important in the
calculation of CP asymmetries in $B\to\pi K$ and $B\to \pi\pi$
decays \cite{6}.

\section{\mbox{\boldmath$K\to\pi\pi$}: strange penguin}

$s\to d$ penguin transition changes the isospin by 1/2 in this way
explaining the famous $\Delta I = 1/2$ rule in $K\to\pi\pi$ decays.
The calculation of the $K_S\to\pi^+ \pi^-$ decay amplitude generated
by the penguin transition using the factorization underestimates the
amplitude by the factor 2-3 according to refs.~\cite{5,7}.

In view of the results for $B$ and $K$ decays we can cautiously
assume that for $D\to \pi^+ \pi^-$ decay the factorization
calculation underestimates the penguin amplitude at most by factor 5 
leading to:
\begin{equation}
\left(\Delta A_{CP}^{\rm theor}\right)_{SM} \la 0.2\% \;\; .
\label{34}
\end{equation}
Thus the following alternative emerges: either  the experimental results are
wrong or New Physics is found. Of course we cannot determine what
kind of new particles and interactions are responsible for large
CPV asymmetry in $D\to \pi^+\pi^-$ ($K^+ K^-$) decays. However, in
the next section we will propose the straightforward
generalization of the Standard Model in which large CPV in $D$
decays can be explained.

\section{The fourth generation: enhancement of CPV in \mbox{\boldmath$D$} decays}

As it was stated in paper \cite{8} the introduction of the fourth
quark-lepton generation may easily remove Standard Model upper
bound (\ref{34}) matching the experimental results \cite{1,2}. In
the case of the fourth generation the additional term with the
intermediate $b'$ quark should be added to the expression for the
penguin amplitude. In this way expression (\ref{4}) is substituted
by:
\begin{eqnarray}
V_{cd} V_{ud}^* f(m_d) + V_{cs}V_{us}^* f(m_s) + V_{cb} V_{ub}^*
f(m_b) + V_{cb'} V_{ub'} f(m_{b'}) = \nonumber \\
= V_{cd} V_{ud}^*[f(m_d)-f(m_s)] + V_{cb}V_{ub}^* [f(m_b)-f(m_s)]
+ V_{cb'} V_{ub'}[f(m_{b'}) -f(m_s)] \;\; , \label{35}
\end{eqnarray}
where the unitarity of 4$\times$4 quark mixing matrix is used.
According to the experimental constraints from the direct searches
of the fourth generation quarks $b'$ should weigh several hundreds
GeV, that is why $f(m_{b'})$ is small and can be neglected just as
it is done with $t$-quark contribution to $b\to s$ penguin, see
the remark after Eq. (\ref{26}). In order to enhance SM
contribution to the penguin amplitude we should suppose that the
term $V_{cb'}V_{ub'}f(m_s)$ dominates.

Then the enhancement of $A_{CP}$ in the case of the fourth
generation is equal to:
\begin{eqnarray}
\frac{P_4}{P_{SM}} & = & \frac{\ln(m_W/m_c)}{\ln(m_b/m_c)}
\frac{|V_{cb'} V_{ub'}^*|}{|V_{cb}V_{ub}|} \frac{\sin(\arg
V_{cb'}V_{ub'}^*)}{\sin\gamma} \approx \nonumber \\
& \approx & 3.3 \frac{3\cdot 10^{-4}}{1.5 \cdot 10^{-4}} \approx 6
\;\; , \label{36}
\end{eqnarray}
where in the last equality we use the allowed
values of the product $|V_{cb'}V_{ub'}^*|\sin(\arg
V_{cb'}V_{ub'})$ taken from Fig. 1 of paper \cite{9} \footnote{Let us
  stress that the logarithmic ($\log\left(m_{W}/m_{c}\right)$)
  enhancement originates not from the diagram with the intermediate
  $b'$ quark but from the term $f(m_{s})$.}. So we see that the
enhancement necessary to describe the experimental data on $\Delta
A_{CP}$ can be achieved in the case of the fourth generation.

\section{Saving baryon number by long-lived fourth generation
neutrino}

If weakly mixed particles exist, then the sphaleron processes can
create the baryon asymmetry of the universe \cite{Dick}. As it is
noted in ref.~\cite{11}, the long-lived fourth generation particles save
baryon asymmetry generated in the early universe from erasure by
the sphaleron transitions. The sphaleron transitions conserve
$B-L$, thus, if in the early universe $B_0 = L_0 \neq 0$ is
generated, then the final baryon and lepton asymmetries being
proportional to $B-L$ are completely erased. If the fourth
generation particles weakly mix with three quark-lepton
generations of the Standard Model, then two additional quantities are
conserved: $B_4 - L_4$ and $L - 3L_4$, where $B_4$ and $L_4$ are
the densities of baryons and leptons of the fourth generation,
while $B$ and $L$ are the densities of baryons and leptons of
three light generations. In ref.~\cite{11} initial asymmetries $B_0 =
L_0 = 3\Delta$ and $B_4^0 = L_4^0 =0$ were chosen and since $L-3L_4 =
3\Delta \neq 0$, the total baryonic number density, $B+B_4$, 
being proportional to a linear superposition of conserved quantities 
is nonzero at the sphaleron
freeze-out temperature. After the sphaleron
freeze-out $B+B_4$ is conserved in comoving volume
and is equal to the present day baryon density of the Universe. 
However, if heavy baryons of the 4th generation do not decay prior to 
big bang nucleosynthesis (BBN), the light baryon number density at BBN could be
different from that determined from the angular fluctuations of CMB. The impact
of this effect  on the light element abundances is discussed below.

For such a scenario to occur the lifetimes of the fourth
generation quarks and leptons should be at least larger than the 
universe age at the sphaleron freeze-out: $\tau_4 > M_{\rm
Pl}/T_{\rm sph}^2 \sim 10^{-10}$ s. For the mixing angles in the case of
$b'\to (c,u)W$ decay it gives $\theta < 10^{-8}$ \cite{11}, much
smaller than what we need to explain the large CPV in $D$-decays, see
eq.~(\ref{36}).

So in our case quarks of the fourth generation should be much stronger
mixed with quarks of three light generations.
However, let us suppose that leptons of the fourth generation
are weakly mixed with the leptons of three light generations. Let
us introduce the total baryon density, $B'\equiv B + B_4$, and take the
initial conditions analogous to those in ref.~\cite{11}: $B'_0 = L_0 =
3\Delta$ and $L_4^0 = 0$.  We can choose four independent chemical potentials
as:  $\mu_{u_L}$, $\mu_W$, $\mu_{N_L}$ and $\mu\equiv
\mu_{\nu_e} + \mu_{\nu_\mu} + \mu_{\nu_\tau}$, which are the
chemical potentials for the upper type quarks, $W$-bosons, $4G$
neutrino and sum over all SM neutrino chemical potentials (see
Appendix). In the limit $\mu_i/T \ll 1$ the baryon and lepton
densities are linear combinations of these chemical potentials
with the coefficients which depend on the ratio of masses of the
corresponding particles to the temperature. We will take into
account the masses of $W$-boson, $t$-quark, $t'$- and $b'$-quarks
of the fourth generation and the fourth generation leptons $N$ and
$E$, the masses of all the other components of the primeval plasma
can be neglected in comparison with $T_{\rm sph}$.

\begin{figure}[ht!]
\begin{center}
\includegraphics[scale=0.4]{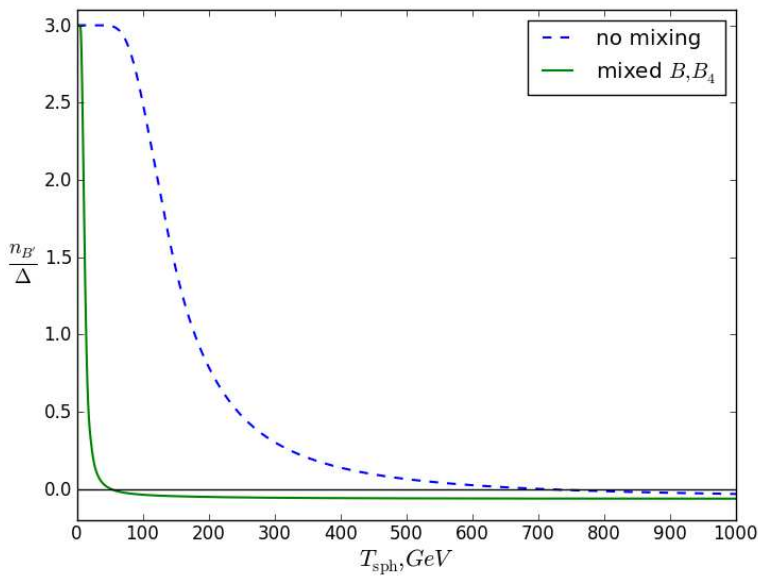}

\caption{The final baryon asymmetry versus the initial asymmetry
$n_{B'}/\Delta$ as a function of sphaleron freeze-out temperature
$T_{\rm sph}$ for the unmixed fourth generation is shown by
a dashed (blue) line. It is analogous to Fig. 2 from \cite{11} but
for $m_N = 57.8$ GeV, $m_E = 107.6$ GeV, $m_{t'} = 634$ GeV,
$m_{b'} = 600$ GeV. The final baryon asymmetry for the case of the
mixed fourth generation quarks and the unmixed fourth generation
leptons is shown by a solid (green) line.}

\end{center}
\end{figure}

Finally we have four equations for four unknown chemical
potentials: two quantities are conserved under the sphaleron
transitions; we can choose them as
\begin{eqnarray}
B' - L - L_4 & = & 0 \;\; , \nonumber \\
L - 3L_4 & = & 3\Delta \;\; . \label{37}
\end{eqnarray}
The third equation is that of the electric neutrality of the primeval plasma,
$Q=0$, and, finally, the sum of the chemical potentials of all the
particles which are converted into nothing by sphaleron ($qqql$ of
each generation) equals zero. The values of masses of the 4th generation
particles we take from paper \cite{12} in which the fit to the
electroweak observables for higgs mass $m_H = 125$ GeV was
performed and recent LHC bounds on the masses of $t'$- and
$b'$-quarks were taken into account:
\begin{eqnarray}
m_{t'} = 634 \; {\rm GeV} & , & \;  m_{b'} = 600 \; {\rm GeV} \;\; ,
\nonumber \\
m_E = 107.6 \; {\rm GeV} & , & \; m_N = 57.8 \; {\rm GeV} \;\; .
\label{38}
\end{eqnarray}

The dashed blue line in Fig. 3 corresponds to the case of the
unmixed fourth generation particles considered in \cite{11}. The
results for the case of the strongly mixed fourth generation
quarks and the unmixed fourth generation leptons are shown by the
solid green line. In order that leptons, $N$, do not decay before the
sphaleron freeze-out, which happens at $t_U \sim 10^{-10}$~s, 
the mixing angles of $N$ with three light neutrinos should be small:
$\theta_i < 10^{-5}$ ($N$ decays through four fermion
interaction). Assuming similar bound $\theta<10^{-6}$, the existence
of heavy Dirac sequential neutrino with $m_{N}=50-100$ GeV is
compatible with the search at LEP II \cite{13}. 

According to the standard cosmological scenario nonrelativistic matter
started to dominate the cosmic energy density at redshift $z \approx 10^4$. 
If we demand that  $N$ should decay before that epoch, its life-time should be 
sufficiently short, $\tau_N < 10^{13}$~s, from which we obtain the lower
bound $\theta> 10^{-16}$. (Let us note that direct searches exclude
$N$ as a unique dark matter candidate \cite{Soni}).

 A stronger bound on $\tau_N$ follows from the equilibrium form of the 
energy spectrum of CMB. According to ref.~\cite{syunyaev} a large influx of
energy into the usual cosmological cosmic background would be thermalized 
if it took place before $z\sim 10^7$. Otherwise the observed black body spectrum of
CMB would be noticeably distorted. Since the precision of the spectral shape is
at the level of $10^{-4}$, only a very small distortion is permitted.

The condition that $N$ decays before or at $z \sim 10^7$ demands 
$\tau_N < 10^6 $ s, or $\theta > 10^{-13}$.  If $N$ indeed decays before
$z\sim 10^7$, the contribution from its decay to the energy density of CMB would be
not larger than 1\% and the ratio of baryon to photon number densities 
$\eta_{B}\equiv n_B /n_\gamma$ at BBN epoch and at CMB recombination would be slightly
different but in principle measurable by the light element abundances.

 More interesting and pronounced effect appears if heavy quarks of the 4th 
generation are long-lived. In this case we cannot explain the large
value of CPV in $D$ decays
but may explain the difference of $\eta_{B}$ at BBN epoch
($\eta_{BBN}$) and at the recombination ($\eta_{rec}$) which is probably requested
by the recent data on the light element abundances~\cite{extra-nu}.  If heavy
baryons of the 4th generation decays after BBN but before the hydrogen 
recombination, the number of light baryons in the comoving volume at BBN 
would be different from that at the recombination. The ratio 
$\eta_{BBN}/\eta_{rec}$ at these epochs could be either larger or
smaller than unity  depending upon the  
value of the baryon asymmetry in the heavy quark sector and the energy influx to CMB
from the heavy baryon decays.  So in principle both rise or decrease of $\eta_{BBN}$ is
possible\footnote{Since both the value of $\eta_{BBN}$ and the number
  of light neutrino species influence nucleosynthesis, the change in
  the value of $\eta_{BBN}$ can be formulated as an additional
  (positive or negative) number of light neutrino species \cite{extra-nu}.}.

In the limit $T_{\rm sph}\to 0$ heavy particles of the fourth
generation are not produced: $B_4 = L_4 =0$, $B' = L = 3\Delta$.
In the physically interesting opposite limit $T_{\rm sph} \gg m_N$
the value of baryon asymmetry is nonzero since the right-handed
neutrinos of three light generations are not produced in the
primordial plasma violating symmetry between the leptons of four
generations which would occur at $T\gg m_N$. The characteristic
time of the right-handed neutrino to thermalize is $T/m_\nu^2$ and
for $m_\nu\la 1$ keV (which is valid for three light neutrinos)
this time is longer than the Universe age,  $t_U = M_{\rm Pl}/T^2$ 
for $T = T_{\rm sph} \approx 200$ GeV, \cite{ Dick}.

\section{Conclusions}

In Introduction we determined what ratio of the penguin to the
tree amplitudes of $D\to\pi^+ \pi^-$ decay is needed to get the
observed CP asymmetry. In Section 2 we found that the
factorization describes the tree amplitude with good accuracy;
concerning the penguin amplitude it appears to be twenty times
smaller than one needs to describe the experimental data on
$A_{CP}$. In Section 3 we demonstrated that in the case of $B\to
\pi^+ K^0$ decay the factorization underestimates the
penguin amplitude by factor 2. In the case of $K_S \to \pi^+
\pi^-$ decay the penguin amplitude is enhanced by factor 2-3 in
comparison with the factorization result.

Thus if confirmed on larger statistics and future systematics 
result (\ref{1}) demands New Physics.

In Section 5 we demonstrated that the fourth quark-lepton
generation may enhance the penguin amplitude describing the
experimental data. If the leptons of the fourth generation weakly
mix with three light generation leptons, then the baryonic charge
generated at high scale escapes the erasure by sphalerons and
survives till now according to the results presented in Section 6.

We are grateful to S.I. Blinnikov for the illuminating discussion
on the chemical potentials, to V.A. Rubakov for the clarifying
discussion on the baryon density in the unbroken electroweak
phase, and to J. Zupan for the remark concerning $D\rightarrow K^{0}\bar
{K^{0}}$ decay.

A.D., S.G., and M.V. acknowledge the support of the grant of the 
Russian Federation government 11.G34.31.0047. S.G. and M.V.
are partially supported by the grants RFBR
11-02-00441, 12-02-00193 and by the grant NSh-3172.2012.2. S.G. is
partially supported by the grant RFBR 10-02-01398.

\bigskip

\begin{center}

{\large \bf Appendix}

\end{center}

\setcounter{equation}{0} \def\theequation{A.\arabic{equation}}

\vspace{3mm}

Below we derive equations used in Section 6 to find the dependence of 
the baryon asymmetry of the Universe on the sphaleron freeze-out 
temperature. In this Appendix we closely follow paper~\cite{11}.

Being interested in the values of the asymmetries at sphaleron
freeze-out temperature we should assume that the electroweak phase
transition already has occured and the neutral Higgs boson condenses.
That is why the Higgs boson chemical potential is zero. Sometimes
in the literature the baryon density in the electroweak unbroken phase
is looked for. In this case the Higgs boson does not condense and
its chemical potential is nonzero. To find it an additional
equation is needed. It is provided by the condition that the
density of charges with which the massless bosons interact should
be zero, and in an unbroken phase there are two such charges: the
hypercharge and the third projection of a weak isospin. The baryon
density in the unbroken phase is analyzed, for example, in  book
\cite{15} and it differs from its value in a broken phase. Since
the right-handed components of quarks and leptons emitting neutral
Higgs transform to the left-handed components the chemical
potential of both components are equal: $\mu_{u_R} = \mu_{u_L} \equiv \mu_u$,
$\mu_{d_R} = \mu_{d_L} \equiv \mu_d$, $\mu_{e_R} = \mu_{e_L}
\equiv \mu_e$. The analogous relations are valid for the particles
of the second and third families. The right-handed neutrinos of
three light generations are not thermalized and should not be
taken into account (see the end of Sect. 6). The fourth generation
right-handed neutrinos, being heavy, rapidly thermalize: $\mu_{N_R}
= \mu_{N_L} \equiv \mu_N$. The chemical potentials of up and down
weak isospin components are related by $W^-$ chemical potential:
$\mu_d = \mu_W + \mu_u$, $\mu_e = \mu_W + \mu_\nu$, $\mu_E = \mu_W
+ \mu_N$. Mixing of quarks of four families and leptons of three
families equilibrates the chemical potentials of the particles with
the identical gauge quantum numbers. As a result 
four independent chemical potentials remain: $\mu_u$, $\mu_N$,
$\mu_W$ and $\mu\equiv \mu_{\nu_1} + \mu_{\nu_2} + \mu_{\nu_3}
\equiv 3\mu_\nu$.

The particle number densities depend on their (Fermi or Bose)
statistics, temperature, chemical potential, and masses.
The chemical potential of an antiparticle is opposite to
that of the particle. The asymmetries and, hence, chemical
potentials are very small. Expanding the equilibrium integrals 
for the asymmetry over $\mu$ we obtain:
$$
n_p = \frac{g_p}{\pi^2}T^3\left(\frac{\mu}{T}\right)
\int\limits_x^\infty y\sqrt{y^2 - x^2} \frac{e^y}{(1\pm e^y)^2}dy
=
$$
\begin{equation}
= \left\{
\begin{array}{ll}
\frac{g_p T^3}{3} \left(\frac{\mu}{T}\right) \alpha_b(x) \; , &
{\rm if} ~ p ~ {\rm is ~ a ~ boson} \\
\frac{g_p T^3}{6} \left(\frac{\mu}{T}\right) \alpha_f(x) \; , &
{\rm if} ~ p ~ {\rm is ~ a ~ fermion} \;\; , \label{A.1}
\end{array}
\right.
\end{equation}
where $g_p$ is the number of the degrees of freedom of the
particle $p$ ($g_q = g_l =2$, $g_\nu =1$, $g_N =2$, $g_W =3$) and
$x=m/T$. Functions $\alpha(x)$ are normalized in such a way
that $\alpha_b(0)=\alpha_f(0) =1$. In what follows we take into
account the nonzero masses of the particles of the fourth
generation, of $t$-quark, and of $W$-boson.

The condition of electroneutrality of the primeval plasma looks as:
\begin{eqnarray}
Q & = & 3\frac{2}{3}[2(\alpha_u + \alpha_c +\alpha_t
+\alpha_{t'})\mu_u] - 3\,\frac{1}{3}[2(\alpha_d + \alpha_s +
\alpha_b +\alpha_{b'}) \times \nonumber \\
&\times &(\mu_W + \mu_u)] -2[(\alpha_e + \alpha_\mu +
\alpha_\tau)(\mu_W + \mu_\nu)] - 2\alpha_E(\mu_W + \mu_N) -
\nonumber \\
& - & 3\cdot 2 \alpha_W \mu_W = 0 \;\; , \label{A.2}
\end{eqnarray}
\begin{equation}
(1+2\alpha_t + 2\alpha_{t'} -\alpha_{b'})\mu_u - (6+\alpha_{b'}
+\alpha_E + 3\alpha_W)\mu_W -\mu - \alpha_E \mu_N = 0. \label{A.3}
\end{equation}
Here and below we omit irrelevant factor $T^{2}/6$.

The sphaleron transition converts $qqql$ combination of each
generation into vacuum, which gives:
\begin{equation}
12\mu_u + 8\mu_W + \mu + \mu_N =0 \;\; . \label{A.4}
\end{equation}

The remaining two equations are two superpositions of $B^\prime$,
$L$, and $L_4$ conserved under sphaleron transitions thus being
equal to their initial values. The expressions for these
quantities look like:
\begin{equation}
L_4 = 2\alpha_E \mu_E + 2\alpha_N \mu_N = 2(\alpha_E + \alpha_N)
\mu_N + 2\alpha_E \mu_W \;\; , \label{A.5}
\end{equation}
\begin{equation}
L = 2(\alpha_e + \alpha_\mu + \alpha_\tau)\mu_e
+(\alpha_{\nu_e}+\alpha_{\nu_\mu} + \alpha_{\nu_\tau})
\frac{\mu}{3} = 3\mu + 6\mu_W \;\; , \label{A.6}
\end{equation}
\begin{eqnarray}
B^\prime & = & 2\cdot 3 \cdot \frac{1}{3}[(\alpha_u + \alpha_c +
\alpha_t +\alpha_{t'})\mu_u +(\alpha_d + \alpha_s + \alpha_b
+\alpha_{b'})\mu_d] = \nonumber \\
& = & 2(2+\alpha_t +\alpha_{t'})\mu_u + 2(3+\alpha_{b'})(\mu_u +
\mu_W) \;\; . \label{A.7}
\end{eqnarray}

Thus we have four equations which determine the chemical
potentials: (\ref{A.3}), (\ref{A.4}), and the remaining two:
\begin{eqnarray}
B^\prime -L -L_4 &=&2(5+\alpha_t +\alpha_{t'} +\alpha_{b'})\mu_u
+2(\alpha_{b'} -\alpha_E)\mu_W -\nonumber\\  &&\hspace{2.55cm}-3\mu -2(\alpha_E +\alpha_N)\mu_N =
0 \;\; , \label{A.8}\\
L - 3L_4 &=& 6(1-\alpha_E)\mu_W + 3\mu -6(\alpha_E +\alpha_N)\mu_N =
3\Delta \;\; , 
\end{eqnarray}

where we take the initial values analogous to those of ref.~\cite{11}:
$B_0^\prime = L_0 = 3\Delta$ and $L_4^0 =0$.

When temperature is much larger than the masses of all the
particles, all $\alpha_i$ are equal to one we obtain:
\begin{equation}
\frac{B^\prime}{\Delta}\left|_{T\gg m_i} = -\frac{11}{179} \right.
\;\; . \label{A.9}
\end{equation}

If the right-handed neutrinos of three light generations
thermalized  then the equation (\ref{A.6}) would be substituted by
\begin{equation}
L=4\mu +6\mu_W \;\; , \label{A.10}
\end{equation}
and the baryon asymmetry at $T\gg m_i$ would vanish.

\end{document}